\documentclass[pra,showpacs]{revtex4}

\usepackage{amsmath}
\usepackage{bm}
\usepackage[dvipdfmx]{graphicx}

\usepackage[dvipdfmx]{color}
\usepackage[dvipdfmx]{graphicx}

\usepackage{float}

\usepackage{ulem}

\begin{document}

\title{Superconducting fluctuation current caused by gravitational drag}

\author{Satoshi Tsuchida} 
\address{Department of Physics, Osaka City University, 3--3--138 Sugimoto, Sumiyoshi--ku, Osaka City, Osaka, 558--8585, Japan
}
\author{Hiroshi Kuratsuji}
\affiliation{Department of Physics, Ritsumeikan University-BKC,  Kusatsu,  525--8577, Japan}



%
\begin{abstract} 

We examine a possible effect of the Lense--Thirring field or gravitational drag by
calculating the fluctuation current through a superconducting ring.
The gravitational drag is induced by a rotating sphere,
on top of which the superconducting ring is placed.
The formulation is based on the Landau-Ginzburg free energy functional of linear form.
The resultant fluctuation current is shown to be 
greatly enhanced in the vicinity of the transition temperature,
and the current also increases on  increasing the winding number of the ring.
These effects would provide a modest step towards magnification of tiny gravity
\end{abstract}

%

\maketitle

{\it{1. Introduction}}:
It is known that there is a close resemblance
between electromagnetic field and weak gravity~\cite{LL1,Moller}.
Specifically the gravitational counterpart of  a magnetic field should be mentioned. 
This is known as the 
Lense--Thirring field~\cite{LL1,Moller,Zee,Ramos,Mash,Ciufolini}, alias gravitational drag, which is generated  near a rotating body.

Half a century ago, DeWitt proposed an idea to detect the quantum effect caused by the gravitational drag by utilizing the characteristics of superconductors~\cite{Dewitt}.
This led to the current caused by the gravito-magnetic potential
arising from a rotating body placed near a superconductor.
Such an attempt belongs to the laboratory experiment of gravity~\cite{Caves},
which is in contrast to the detection of the quantum effects
caused by the Earth gravity using, e.g.,  a neutron interferometer~\cite{Green,Collera}. 
The use of a superconductor has an advantage;
huge enhancement of the current would be a possibility 
by using the coherence nature characterizing the superconductivity.
The study of gravitation using the superconductivity has been explored as
a specific category~\cite{Chiao,Tajmar,Anandan}.

In this note, we address a problem belonging o the same category as Ref.~\cite{Dewitt},
which  deals with the effects occurring in superconductors caused by the gravitational drag.  
However, apart from the DeWitt attempt, we are concerned with an estimation of {\it fluctuation current}  occurring
 in  superconducting ring such that it is arranged in a critical condition~\cite{Schmid,Imry,Langer}.

\begin{figure}[b]
  \begin{center}
    \includegraphics[width=50mm]{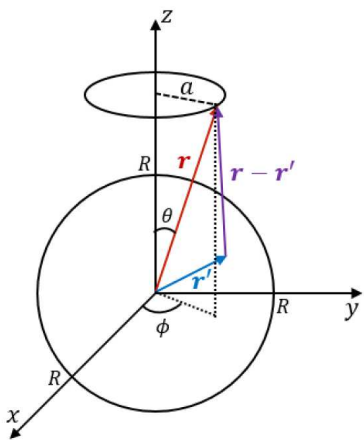}
    \caption{(Color online)
      The setting for detecting the current caused by the gravitational drag induced by the rotating body.
      The coordinates $ \bf{r} $ is the observational point,
      and $ \bf{r}' $ is a point in the sphere to be integrated.
    }
    \label{fig:drag1}
  \end{center}
\end{figure}

{\it{2. Gravitational drag}}:
Let us consider the setting for detecting the current caused by
the gravitational drag due to a rotating body,
which is kept to be electrically neutral.
The body  is assumed to form a spherical shape with radius $ R $,
the mass density $ \rho $, and the total mass $ M $.
The sphere rotates with the angular velocity $ \omega $.
The induced gravitational field is calculated by 
the {\it Amp{\'e}re} law~\cite{Zee}:  
\begin{eqnarray*}
  \nabla  \times {\bf B}_g = - \frac{ 16 \pi G }{ c^{3} } {\bf J} 
\end{eqnarray*}
for the gravitomagnetic field $ {\bf B}_g = \nabla \times {\bf A}_g $
where $ {\bf{A}}_{g} $
means the corresponding vector potential.
$ G $ represents the gravitational constant and
$ {\bf J} $ is the mass current: $ {\bf J} = \rho {\bf v} $ with the velocity $ {\bf v} = {\bf{\omega}} \times {\bf r} $. 
Using the gauge condition: $ \nabla \cdot {\bf A}_g = 0 $, 
it follows that 
$$  \nabla^2 {\bf A}_g = \frac{ 16 \pi G }{ c^{3} } {\bf J} 
$$
which is solved as 
\begin{equation}
{\bf A}_g = - \frac{ 4 G }{ c^{3} } \int \frac{{\bf J({\bf r}')}}{\vert {\bf r}- {\bf r}'\vert}d{\bf r}' 
\end{equation}
The coordinates $ \bf{r} $ and $ {\bf{r}}' $ are given in Fig.~\ref{fig:drag1}.
By using the multipole expansion, this is calculated as
\cite{LL1,Ramos}
\begin{eqnarray}
  {\bf{A}}_g = - \frac{ 2 G }{ c^{3} r^{2} } {\bf{M}} \times {\bf{n}}
\end{eqnarray}
where $ {\bf{n}}$ represents a unit vector direction of $ {\bf{r}} $ and $ {\bf{M}} $ is angular momentum of rotating sphere.
The vector potential $ {\bf{A}}_{g} $ is alternatively written
in term of angle $ \theta $ and $ \phi $, which are indicated in Fig.~\ref{fig:drag1}, as 
\begin{eqnarray}
  {\bf{A}}_g= \frac{ 4 G M \omega R^{2} }{ 5 c^{3} r^{2} } \left( \sin \theta \sin \phi~{\bf{e}}_{x} - \sin \theta \cos \phi~{\bf{e}}_{y} \right)
\end{eqnarray}
where $ {\bf{e}}_{x} $ and $ {\bf{e}}_{y} $ are the unit vectors direction of $ x $ and $ y $,
respectively. \\

{\it{3. Fluctuating current}}:
Let us consider a superconducting circular ring, which is placed above the rotating body such that 
the ring is coaxial to the sphere, and the radius is given by $ a $ which is prescribed to be smaller than 
the radius of the sphere as shown in Fig.~\ref{fig:drag1}. 
We adopt an ideal limit; the thickness of the ring is neglected.
That is, the radius $ a $ is extremely larger than the cross section of the ring.

Our argument is based on the Landau--Ginzburg (LG) free energy, for the order parameter $ \psi $, 
that is written as~\cite{Schmid,Imry,Langer,LL2}
\begin{eqnarray*}
 F_{g} = \int \left[ \frac{1}{2m} \psi^{*} \left( -i \hbar \nabla - m c {\bf A}_{g} \right)^2 \psi + V( \psi^{*} , \psi ) \right] d {\boldsymbol{x}}  
\end{eqnarray*}
and the corresponding partition function is given by the functional integral:
\begin{equation}
  \label{eq:partfunc}
  Z = \int \exp[ -\beta F_g] \mathcal{D}[\psi, \psi^{*}]
\end{equation}
We make the following remark:
We are concerned with the fluctuation current in the vicinity just above the 
superconducting transition temperature.
Hence, consideration of  the Meissner effect is not relevant in contrast to DeWitt procedure \cite{Dewitt}, 
so no current is induced as a reaction of the Meissner effect in the gravito-magnetic field. 
Thus the gravito-magnetic vector potential directly enters in the same form as the usual electromagnetic one. 
As is implied from the above prescription, 
the gravito-magnetic vector potential $ A_{g} $ has only $ \phi $ component
in terms of polar coordinate, which relates to electromagnetic vector potential $ A $ as follows:
\begin{eqnarray}
  A = - \frac{ mc^{2} }{ 2e } A_{g}
\end{eqnarray}
In the circular geometry, we have $ F_{g} = \int \psi^{*}\Lambda \psi d \phi $,
where $ \Lambda $ is given by 
\begin{eqnarray}
  \Lambda = \frac{ \hbar^{2} }{ 2 m a^{2} } \left( -i \frac{ d }{ d \phi } + \frac{\Phi}{\phi_0} \right)^{2} 
  + \vert \alpha \vert,
  \label{eq:lambdahami}
\end{eqnarray}
with the flux $ \Phi = \int A a d \phi $,
and the flux quantum $ \phi_0 = hc/2e $ ($ h $ means the Planck constant). 
Here, we note that
the flux $ \Phi $ is multiplied by $ N $ when the winding number of the ring is $ N $ times that of a basic circular ring.
Equation~(\ref{eq:lambdahami}) is the linear LG theory, for which the potential term $ V $ 
is $ \alpha \psi^{*}\psi $ and the parameter $ \alpha $ is regarded as the ``mass'',
which is assumed to take a form $ \alpha = C ( T - T_{c} ) $
($ C $ is some constant, and $ T_{c} $ means the transition temperature)~\cite{Schmid,Imry,LL2}. 
The quartic term is omitted since the fluctuation effect is
very small near the transition temperature.

Let us expand the order parameter as $ \psi (t) = \sum_{n} a_{n} (t) \psi_{n} $
with $ n $ taking an  integer value running from $ - \infty $ to $ + \infty $ . 
Here $ \psi_{n} $ are the eigenfunctions for the eigenvalue equations: $ \Lambda \psi_{n} = \lambda_{n} \psi_{n} $.
The eigenvalues are calculated to be
\begin{eqnarray}
  \lambda_{n} = \frac{ {\hbar}^{2} }{ 2 m a^{2} } \left( n + \frac{ \Phi}{ \phi_{0} } \right)^{2} + \vert \alpha \vert
  \label{eq:smalllambda}
\end{eqnarray}
and the corresponding eigenfunctions become $ \psi_{n} = \frac{ 1 }{ \sqrt{2 \pi} } \exp \left[ i n \phi \right] $.
The partition function is thus calculated by using the functional integral~(\ref{eq:partfunc}):
\begin{eqnarray}
  Z &=& \int \exp \left[ - \beta \sum_{n} \lambda_{n} a_{n}^{*} a_{n} \right] \prod_{n} da_{n}^{*} da_{n} \nonumber \\
  &=& \prod_{ n = - \infty }^{ + \infty } \frac{ 2 \pi }{ \beta \lambda_{n} }
  \label{partition}
\end{eqnarray}
from which one obtains the free energy $ F = - k_{B} T \log Z $ 
(up to some additional constant).
This is calculated as 
\begin{eqnarray}
  F &=& k_{B} T \sum_{n} \log \left[ \left( n + \frac{ \Phi }{ \phi_{0} } \right)^{2} + \vert \gamma \vert \right] \nonumber \\
  &=& k_{B} T \log \left[ \cosh ( 2 \pi \sqrt{\gamma} ) - \cos \left( 2 \pi \frac{ \Phi }{ \phi_{0} } \right) \right]
\end{eqnarray}
Here we have used the formula \cite{math}\footnote{This formula was suggested by Mr. Hideki Ono; 
the former graduate student of one of the present authors H.K..}

\begin{eqnarray}
  \sum_{n = -\infty }^{ +\infty } \frac{ 1 }{\left( n + \frac{ \Phi }{ \phi_{0} } \right)^{2} + x^{2} }
          = \frac{ \pi }{ x } \frac{ \sinh( 2 \pi x ) }{ \cosh ( 2 \pi x ) - \cos \left( 2 \pi \frac{ \Phi }{ \phi_{0} } \right) } 
  \label{sum}
\end{eqnarray}
Hence we can derive the
{\it fluctuating current} along the ring, which is obtained by the formula:
$ J =  c \frac{ \partial F }{ \partial \Phi } $, i.e., 
\begin{eqnarray}
  J = \frac{ 2 \pi c k_{B} T }{ \phi_{0} } \frac{ \sin \left( 2 \pi \frac{ \Phi }{ \phi_{0} } \right) }
        { \cosh( 2 \pi \sqrt{\gamma} ) - \cos \left( 2 \pi \frac{ \Phi }{ \phi_{0} } \right) }
\end{eqnarray}
and use is made of the notation 
\begin{equation}  
  \gamma = \frac{ 2 m a^{2} }{ {\hbar}^{2} } \alpha  
  \label{gamma}
\end{equation}
%
together with the flux: 
\begin{eqnarray}
  \left| \Phi \right| &=& 2 \pi a \times \frac{ m c^{2} }{ 2 e } \times \left| {\boldsymbol{A}}_{g} \right| \nonumber \\
       &=& \frac{ 4 \pi m }{ 5 e c } \times \frac{ G M V R a }{ r^{2} }
\end{eqnarray}
where $ V $ represents the rotating velocity $ V = R\omega $. 

If we are concerned with the state that is very close to the transition temperature; $ T= T_C $, 
$ \gamma $ can be regarded as a small quantity by noting $ \alpha = C(T-T_C) $. 
Further $ \Phi/\phi_0 $ is quite a small quantity by taking account of the fact that 
$ G $ and  $ m/e $ are small in absolute sense.
In addition, the second order of $ \Phi/\phi_0 $ is negligibly small,
because the second order is smaller than the first order by 20 orders  of magnitude.
Under this condition,  the current may be approximated as\footnote{We use the relation: 
$ \sin \left( 2\pi \Phi/\phi_0 \right) \simeq 2 \pi {\Phi} / {\phi_{0} } ,~\cos \left( 2 \pi \Phi / \phi_0 \right) \simeq 1,~\cosh \left( 2 \pi \sqrt \gamma \right) 
\simeq  1 + \frac{1}{2} \left( 2\pi \sqrt\gamma \right)^{2}. $}
\begin{equation}
  J \sim 2 c k_{B} T \frac{ \Phi }{ {\phi}_{0}^{2} \gamma }  
\end{equation}
The fluctuating current can be detected by observing (real)  magnetic field produced by the Amp{\'e}re law, which 
directs along the $z$-axis through the center of the circle,  and the magnitude is $ B = \frac{J}{a} $.  \\

{\it{4. Discussion }}:
In order to examine the behavior of the current in the very vicinity of the critical temperature, 
we rewrite this by introducing the  ``coherence length'', $ \xi = \frac{ \hbar }{ \sqrt{ 2 m \left| \alpha \right|} }$ \cite{Schmid}, 
which is sufficiently  larger than  the size of the ring $ a $, namely, 
\begin{eqnarray}
  J = \frac{8}{5} \frac{ k_{B} T }{ \hbar c } \frac{ m e }{ \hbar c } \frac{ G M V R }{ r^{2} a } {\xi}^{2}(T) 
  \label{current}
\end{eqnarray}
Near $ T = T_C $, it is known that $  \xi^2(T) = \xi_0^2 \left( \frac{T_C}{T-T_C} \right)  $ \cite{Schmid,Imry},   
where $ \xi_0 $ means the coherence length at zero temperature, (which may be chosen to be the order $ 10^{-5}~{\rm m} $). By noting this 
fact, the current  expression (\ref{current}) indicates  that the current gets a huge enhancement near $ T = T_{c} $. 
Let us examine the order estimate by assuming 
the freely chosen parameters; $ T,M,V,R,r,a $ as  
\begin{eqnarray}
  J \sim 10^{-20} {\xi}^{2}(T)\left( \frac{ T }{ 10~{\rm{K}} } \right) \left( \frac{ M }{ 10^{3}~{\rm{kg}} } \right) \left( \frac{ V }{ 10^{4}~{\rm{m/s}} } \right) 
 & & \hfill \nonumber \\
  \left( \frac{ R }{ 2~{\rm{m}} } \right)  
\left( \frac{ 2~{\rm{m}} }{ r } \right)^{2}
  \left( \frac{ 10^{-3}~{\rm{m}} }{ a } \right)~[{\rm{A}}]
\end{eqnarray}
If we could control the temperature $ T $ such that it is extremely close to the transition temperature, we expect  $ \xi^2(T) \sim 10^{-2} $, which leads to 
the current of the order $ 10^{-22} $. 
In addition to the critical behavior, the flux $ \Phi $ is proportional to winding number $ N $ of the ring,
so the current $ J $ is also amplified  by $ N $ times.  If we could take $ N \sim 10^5 $, 
the resultant current would be  $ J \sim 10^{-22} \times 10^5 = 10^{-17}[A]  $. 

Thus the present result is quite different from the classic DeWitt's one\cite{Dewitt}: the latter does not incorporate any enhancement mechanism arising 
from the critical temperature and the winding number.  
The enhancement of the present model is consequence of the LG formalism which is concerned 
with the state near critical temperature.   As such,  the order estimate given above is still  tiny, even if  it is magnified. 

Despite such a feature,  the present approach may open a novel aspect of exploring a possible effect  of gravitational drag, which is 
provided by a framework of the LG theory.   The detectability of this effect will depend on experimental  innovations in the future. \\

This article appeared in {\bf Progress of Theoretical and Experimental Physics}, 2017, 12I101 DOI: 10.1093/ptep/ptx171.

\end{document}